\documentstyle[12pt,preprint]{aastex}  

\begin{document}

\title{The Color Differences of Kuiper Belt Objects in Resonance with Neptune}
\author{Scott S. Sheppard\altaffilmark{1}}

\altaffiltext{1}{Department of Terrestrial Magnetism, Carnegie Institution of Washington, 5241 Broad Branch Rd. NW, Washington, DC 20015, USA, sheppard@dtm.ciw.edu}

\begin{abstract}  

The optical colors of 58 objects in mean motion resonance with Neptune
were obtained.  The various Neptune resonant populations were found to
have significantly different surface color distributions.  The 5:3 and
7:4 resonances have semi-major axes near the middle of the main Kuiper
Belt and both are dominated by ultra-red material (spectral gradient:
$S\gtrsim 25$).  The 5:3 and 7:4 resonances have statistically the
same color distribution as the low inclination ``cold'' classical
belt.  The inner 4:3 and distant 5:2 resonances have objects with
mostly moderately red colors ($S\sim 15$), similar to the scattered
and detached disk populations.  The 2:1 resonance, which is near the
outer edge of the main Kuiper Belt, has a large range of colors with
similar numbers of moderately red and ultra-red objects at all
inclinations.  The 2:1 resonance was also found to have a very rare
neutral colored object showing that the 2:1 resonance is really a mix
of all object types.  The inner 3:2 resonance, like the outer 2:1, has
a large range of objects from neutral to ultra-red.  The Neptune
Trojans (1:1 resonance) are only slightly red ($S\sim 9$), similar to
the Jupiter Trojans.  The inner 5:4 resonance only has four objects
with measured colors but shows equal numbers of ultra-red and
moderately red objects.  The 9:5, 12:5, 7:3, 3:1 and 11:3 resonances
do not have reliable color distribution statistics since few objects
have been observed in these resonances, though it appears noteworthy
that all three of the measured 3:1 objects have only moderately red
colors, similar to the 4:3 and 5:2 resonances.  The different color
distributions of objects in mean motion resonance with Neptune are
likely a result from the disruption of the primordial Kuiper Belt from
the scattering and migration of the giant planets.  The few low
inclination objects known in the outer 2:1 and 5:2 resonances are
mostly only moderately red.  This suggests if the 2:1 and 5:2 have a
cold low inclination component, the objects likely had a significantly
different origin than the ultra-red dominated cold components of the
cold classical belt and 5:3 and 7:4 resonances.

\end{abstract}


\section{Introduction}

There is still a debate as to where the objects in the Kuiper Belt
came from.  It is likely that objects that originally formed in the
giant planet region currently reside in the Kuiper Belt along with
objects that formed beyond the giant planets (Gomes et al. 2003;
Levison and Morbidelli 2003; Levison et al. 2008; Walsh et al. 2011).
These various origins for the Kuiper Belt objects (KBOs) are one of
the explanations as to why the colors of the KBOs have been found to
be so diverse (Luu and Jewitt 1996).  The environmental conditions
experienced by KBOs such as space weathering, cratering and
fragmentation may also cause the surfaces of the KBOs to change over
time.  Thus the surface color of a particular KBO is likely a
combination of its original formation location within the solar nebula
and the environmental conditions the KBO has experienced over the age
of the solar system.

Dynamically there appear to be three main types of KBOs.  (1) Objects
that have their perihelion near Neptune ($q \sim 25 - 35$ AU) and have
large eccentricities ($e>0.4$) are called scattered disk objects.  The
scattered disk was likely created through KBOs having strong dynamical
interactions with Neptune (Duncan and Levison 1997; Duncan 2008; Gomes
et al. 2008).  The detached disk objects, like the scattered disk,
have moderate to large eccentricities ($e>0.25$) but with higher
perihelia ($q\gtrsim 40$ AU) and are thus unlikely to have been
scattered by Neptune in the current solar system configuration.
Detached disk objects are likely just fossilized scattered disk
objects from the time when Neptune was still migrating outwards
(Gladman et al. 2002; Lykawka and Mukai 2006; Gomes et al. 2011).  (2)
Objects with semi-major axes $42 \lesssim a \lesssim 48$ AU with
moderate to low eccentricities are considered to be in the main Kuiper
Belt and are called classical KBOs.  The classical objects are usually
split into two sub-categories based on inclination (Brown 2001).
Objects with inclination less than 5 to 10 degrees are called low
inclination or ``cold'' classical KBOs while those with higher
inclinations are considered the ``hot'' classical KBOs.  The low
inclination ``cold'' classicals are redder, smaller and more prevalent
in equal-sized binaries than other populations (Tegler and Romanishin
2000; Levison and Stern 2001; Trujillo and Brown 2002; Stern 2002;
Gulbis et al. 2006; Noll et al. 2008; Peixinho et al. 2008).  Tegler
and Romanishin (2000) suggest that redder objects are based on
eccentricity and perihelion distance and not simply inclination, with
redder objects having larger eccentricities and perihelia.  We find
there is no significant canonical hot classical population as almost
all objects with inclinations above 10 degrees in the classical region
have much larger eccentricities and lower perihelia than the cold
classicals.  Thus the standard hot classical population is more like
the scattered disk objects than classical objects.  The cold classical
population likely formed relatively nearby their current locations as
indicative of their low inclinations and eccentricities.  In contrast,
the hot and scattered population were likely scattered and captured
into their current orbits based on their highly disturbed inclinations
and eccentricities (Batygin et al. 2011; Wolff et al. 2012; Dawson and
Murray-Clay 2012).

\begin{figure}
\epsscale{0.4}
\centerline{\includegraphics[angle=90,totalheight=0.6\textheight]{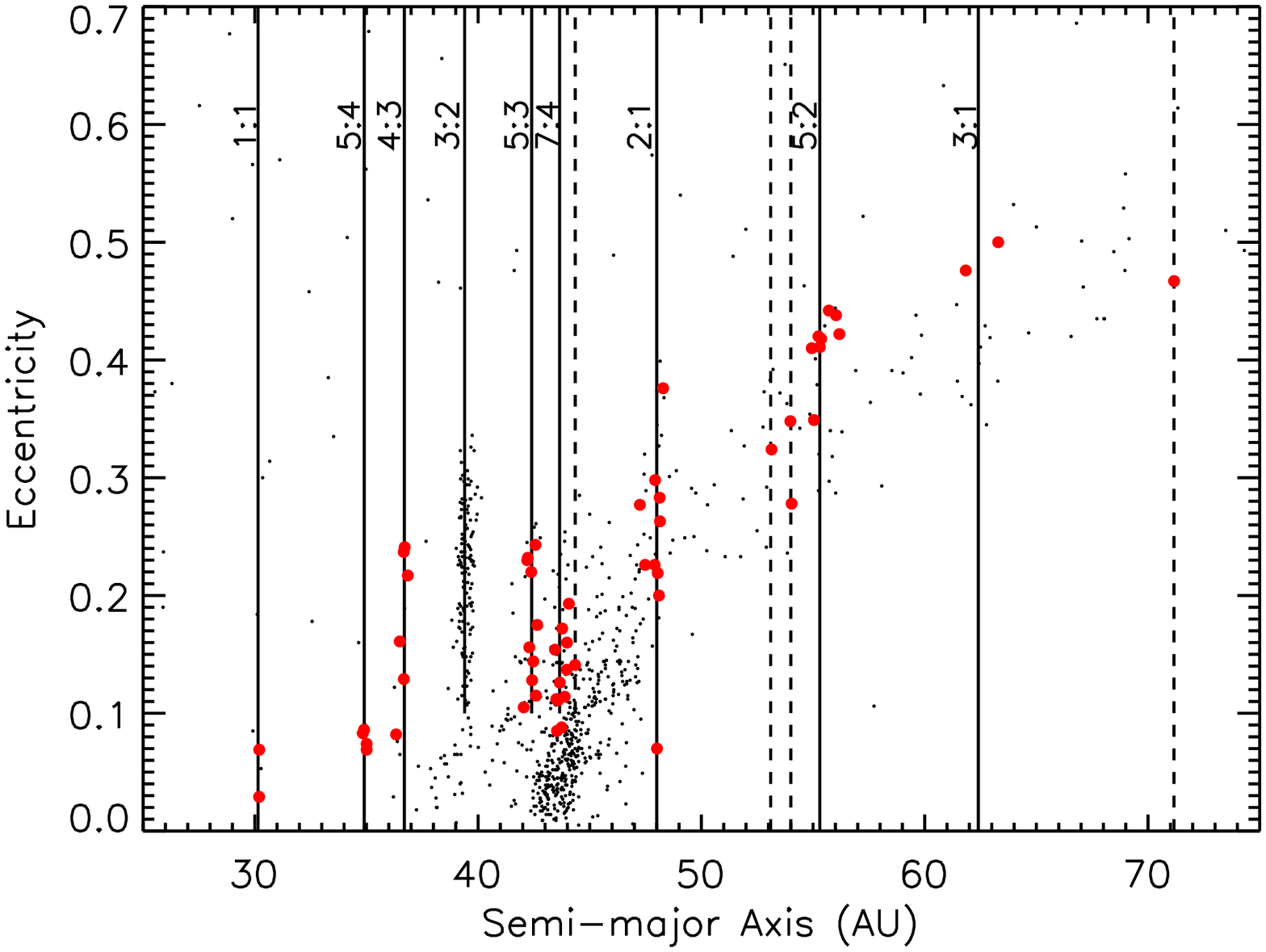}}
\caption{The semi-major axis versus eccentricity for all
  multi-opposition observed TNOs.  Objects observed in this work are
  shown with big filled red circles.  The main Neptune mean motion
  resonances are shown by vertical solid lines.  The less important
  higher order resonances in which objects were observed in this work
  are shown with dashed lines (9:5, 7:3, 12:5, 11:3).}
\label{fig:kboeares} 
\end{figure}

\begin{figure}
\epsscale{0.4}
\centerline{\includegraphics[angle=90,totalheight=0.6\textheight]{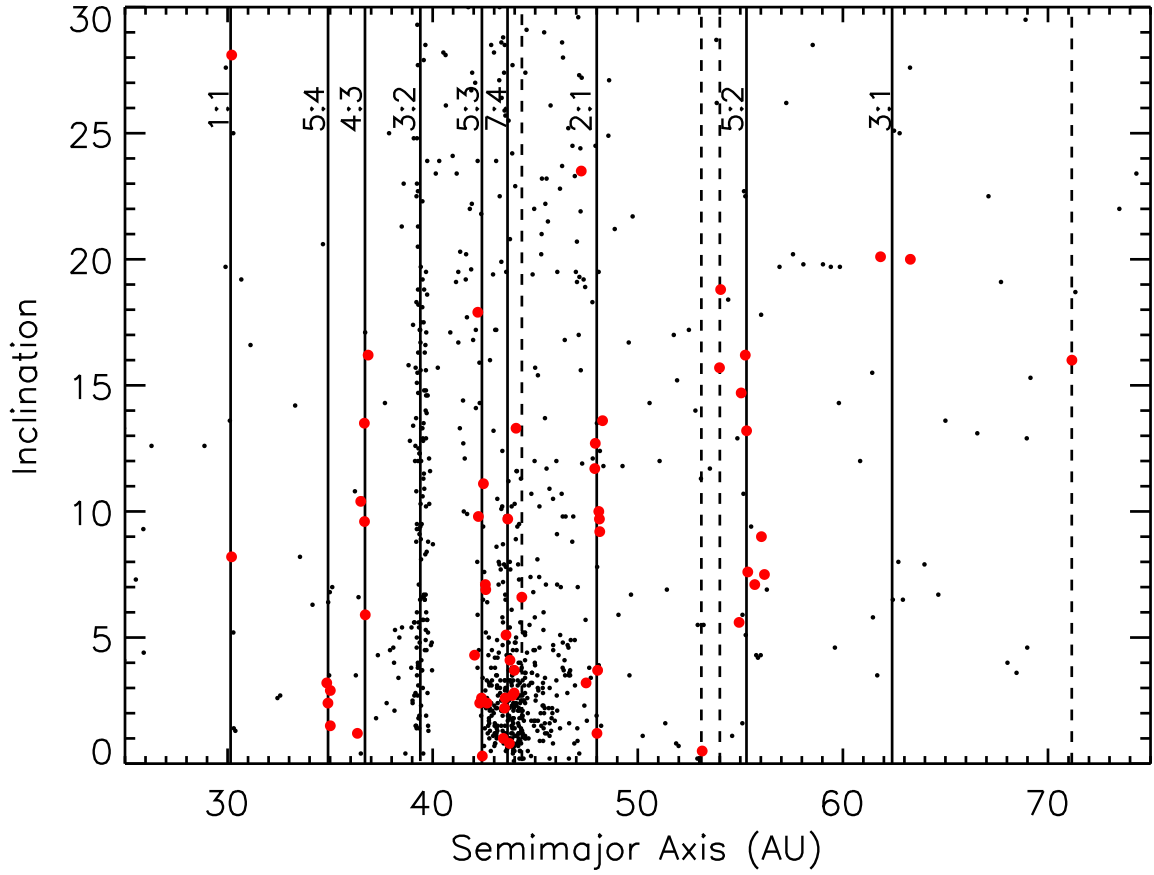}}
\caption{The semi-major axis versus inclination for all
  multi-opposition observed TNOs.  Objects observed in this work
  are shown with big filled red circles.}
\label{fig:kboiares} 
\end{figure}

(3) Resonant KBOs, the main focus of this work, are objects that are
in mean motion resonance with Neptune (Figures~\ref{fig:kboeares}
and~\ref{fig:kboiares}: see Elliot et al. 2005, Lykawka and Mukai 2007
and Gladman et al. 2008 for definitions of the various mean motion
resonances with Neptune and
www.boulder.swri.edu/$\sim$buie/kbo/kbofollowup.html for an updated list of
Neptune resonant objects kept by Marc Buie).  Resonant objects were
likely captured into their respective resonance from the outward
migration of Neptune in the early solar system (Malhotra 1995; Chiang
and Jordan 2002; Chiang et al. 2003; Hahn and Malhotra 2005;
Murray-Clay and Chiang 2005; Levison et al. 2008).  The resonances
with sizable known populations are the Neptune Trojans (1:1), the
resonances with semi-major axes interior to the main classical Kuiper
Belt 5:4, 4:3 and 3:2 (called Plutinos because Pluto is in this
resonance), the 5:3 and 7:4 which have semi-major axes within the main
classical Kuiper Belt and the outer resonances with semi-major axes
exterior to the main classical Kuiper Belt 2:1 (called the Twotinos),
7:3, 5:2, and 3:1.

Where the resonant objects originated and how the resonant objects
came to reside where they are today is still unknown.  The dynamical
and physical properties of KBOs in resonance with Neptune are
extremely valuable since these objects were likely captured into these
resonances from the outward migration of Neptune.  The various orbital
and physical characteristics of the resonant objects will help
constrain the migration and evolution of the planets.  It is likely
that the different resonances swept up or captured objects from
different initial locations.  One way to try to understand how the
resonances became populated is to determine the physical
characteristics of individual resonance objects and compare them to
other populations of small solar system objects.

Were the ultra-red objects emplaced into the Kuiper Belt or did they
form in-situ?  One of the simplest ways to try and answer this
question is to catalog the current locations of ultra-red objects
($S\gtrsim 25$: Jewitt 2002; Sheppard 2010).  Ultra-red material is
mostly associated with the dynamcially stable cold classical KBOs and
possibly the Oort cloud (Tegler and Romanishin 2000; Trujillo and
Brown 2002; Peixinho et al. 2008; Sheppard 2010).  Ultra-red colors
are likely created from material rich in very volatile ices and
organics and is mostly only seen on objects kept far from the Sun
(Jewitt 2002; Grundy 2009; Sheppard 2010; Brown et al. 2011; Merlin et
al. 2012).

This work observed the various KBOs in mean motion resonance with
Neptune for their surface colors in order to look for similarities and
differences between the objects in resonances and other classes of
small solar system objects.  The 3:2 resonance objects have been well
explored physically because they are relatively brighter objects as
they are located near the inner edge of the Kuiper Belt.  Thus past
color data of the 3:2 resonance objects is used and the focus of the
new observations presented in this work is on the colors of the
objects in the little explored more distant heavily populated
resonances such as the 5:4, 4:3, 5:3, 7:4, 2:1, 5:2 and 3:1 as well as
a few lesser populated resonances and Neptune Trojans in the 1:1
resonance with Neptune.

\section{Observations}

All new Kuiper Belt object color measurements presented in this work
are from observations using the Magellan 6.5 meter telescopes at Las
Campanas, Chile.  Table 1 shows the geometry of the observations for
the 58 objects observed.  One of three imaging cameras were used
during the observations with a standard set of Sloan filters.  All
observations were calibrated to Southern Sloan standard star fields
G158-100, PG1633+099 or DLS-1359-11 (Smith et al. 2005).  The LDSS3
imager on the Magellan-Clay telescope was used on the nights of 26
August 2009, and 20 and 21 March 2010.  LDSS3 is a CCD imager with one
STA0500A $4064\times4064$ CCD and $15\micron$ pixels.  The field of
view is about 8.3 arcminutes in diameter with a scale of $0.189$
arcseconds per pixel.  The IMACS camera on the Magellan-Baade
telescope was used on the nights of 20-21 April, 18 August, 11-12
September 2010, 27-28 September 2011 and 23-25 March 2012.  IMACS is a
wide-field CCD imager that has eight $2048\times4096$ pixel CCDs with
a pixel scale of $0.20$ arcseconds per pixel.  The eight CCDs are
arranged in a box pattern with four above and four below and about 12
arcsecond gaps between chips.  Only chip 2 of IMACS, which is just
North and West of the camera center, was used in this analysis.  The
MegaCam imager on the Magellan-Clay telescope was used on the nights
of 9-10 October 2010.  MegaCam has 36 CCDs of $2048\times 4608$ pixels
each with a pixel scale of 0.08 arcseconds per pixel giving a total
field-of-view of $24\arcmin \times 24\arcmin$.  Only chip 23, near the
center, was used in this analysis.

\begin{figure}
\epsscale{0.4}
\centerline{\includegraphics[angle=90,totalheight=0.6\textheight]{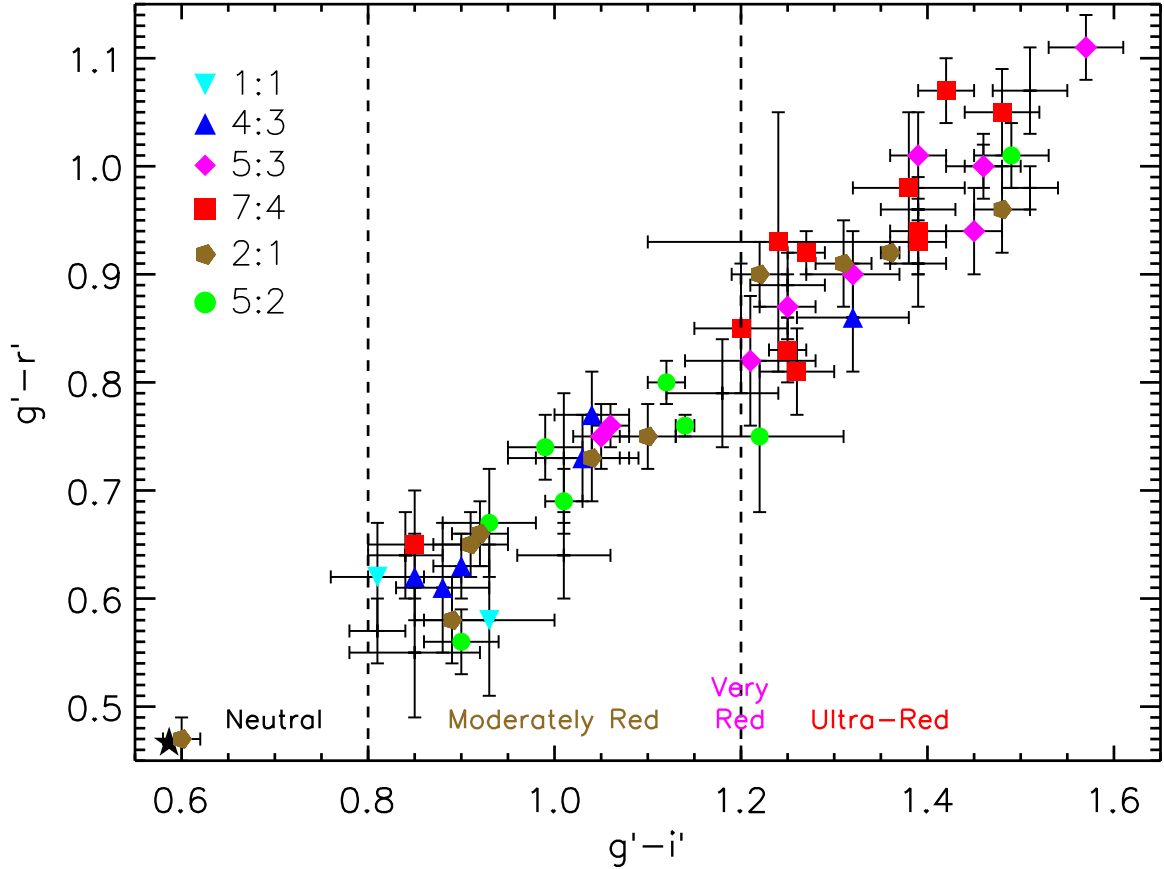}}
\caption{New Sloan g',r' and i' colors for Kuiper Belt objects observed in
  this work.  The middle distance 5:3 (purple diamonds) and 7:4 (red
  squares) resonance populations appear mostly in the ultra-red
  portion of the figure (upper right) and are similar to the color of
  low inclination Classical Kuiper Belt objects.  The inner 4:3 (blue
  triangles) and distant 5:2 (green circles) resonance populations are
  mostly only moderately red and similar to the colors of the comets,
  Jupiter Trojans and Neptune Trojans.  The distant 2:1 appears to be
  a mix of neutral, moderately red and ultra-red objects (brown
  pentagons).  Objects observed in this work in other resonances other
  than the main ones listed above are shown by plus signs.  For
  reference the color of the Sun is marked by a filled black star.
  The ultra-red color is only seen on outer solar system objects and
  is defined here as the color $75\%$ of the cold classical belt
  objects have.  Very-red is defined as the color $90\%$ of the cold
  classical belt objects have.}
\label{fig:RescolorallSloan}
\end{figure}

The data analysis was done in the same way as described in Sheppard
(2010).  Biases and dithered twilight and dome flats were used to
reduce each image.  Images were obtained through either the Sloan g',
r' or i' filter while the telescope was auto-guiding at sidereal rates
using a nearby bright star.  Exposure times were between 300 and 350
seconds.  Filters were rotated after each observation to prevent a
light curve from influencing the color calculations.  To be able to
directly compare our results with previous works the Sloan colors were
converted to the Johnson-Morgan-Cousins BVRI color system using
transfer equations from Smith et al. (2002): $B=g'+0.47(g'-r') +0.17$;
$V=g'-0.55(g'-r')-0.03$; $V-R=0.59(g'-r')+0.11$;
$R-I=1.00(r'-i')+0.21$.  These transformation equations from Sloan to
the BVRI color system were shown in Sheppard (2010) to be good to
within a hundredth of a magnitude by observing bright TNOs with well
known BVRI colors with the Sloan filters.  The Sloan photometric
observations are shown in Table 2 (Figure~\ref{fig:RescolorallSloan})
and the BVRI derived photometric results are shown in Table 3
(Figure~\ref{fig:RescolorallBVRI}) with the orbital information of the
objects given in Table 4.

Photometry was performed by optimizing the signal-to-noise ratio of
the faint small outer Solar System objects.  Aperture correction
photometry was done by using a small aperture on the TNOs ($0.\arcsec
64$ to $1\arcsec$ in radius) and both the same small aperture and a
large aperture ($2.\arcsec 24$ to $3. \arcsec 6$ in radius) on several
nearby unsaturated bright field stars with similar Point Spread
Functions (PSFs).  The magnitude within the small aperture used for
the TNOs was corrected by determining the correction from the small to
the large aperture using the PSF of the field stars (cf. Tegler and
Romanishin 2000; Jewitt and Luu 2001; Sheppard 2010).  

\section{Results and Discussion}

\begin{figure}
\epsscale{0.4}
\centerline{\includegraphics[angle=90,totalheight=0.6\textheight]{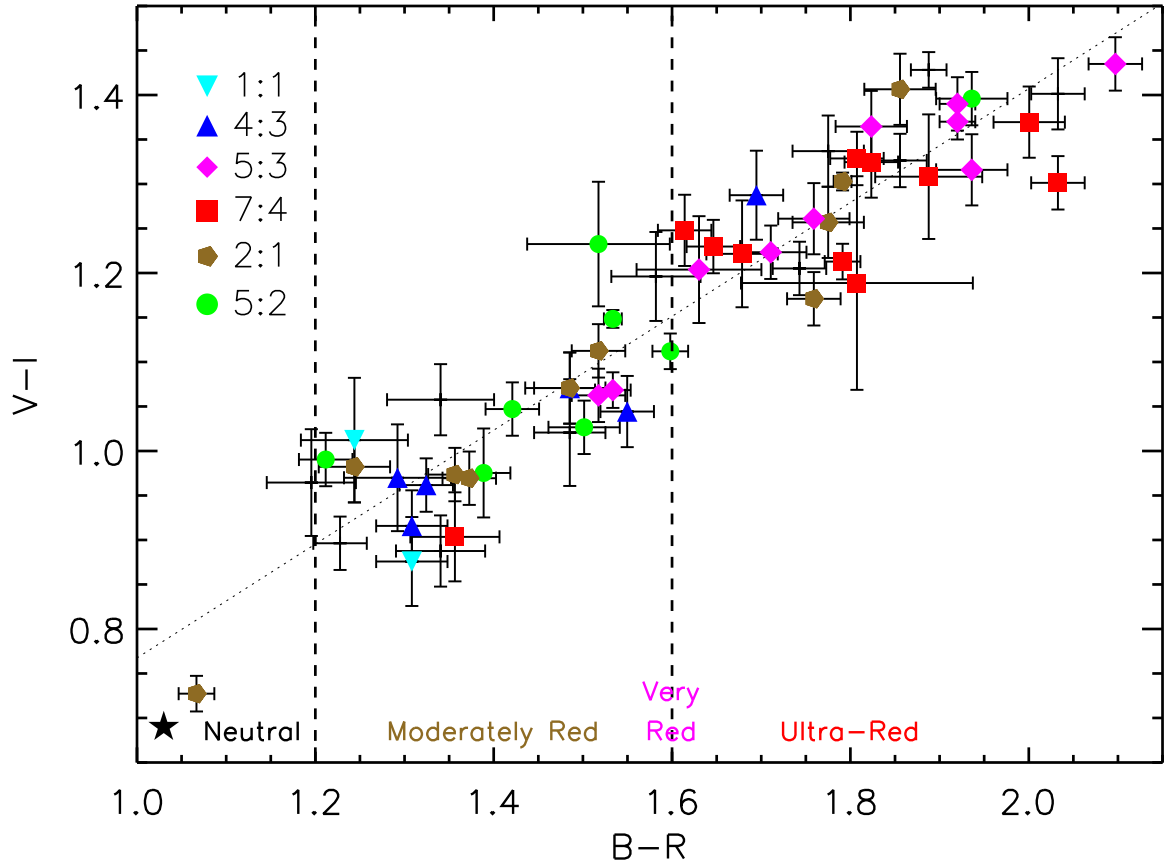}}
\caption{Same as Figure~\ref{fig:RescolorallSloan} except for BVRI
  colors B-R and V-I for objects observed in this work.}
\label{fig:RescolorallBVRI}
\end{figure}

\begin{figure}
\epsscale{0.4}
\centerline{\includegraphics[angle=90,totalheight=0.6\textheight]{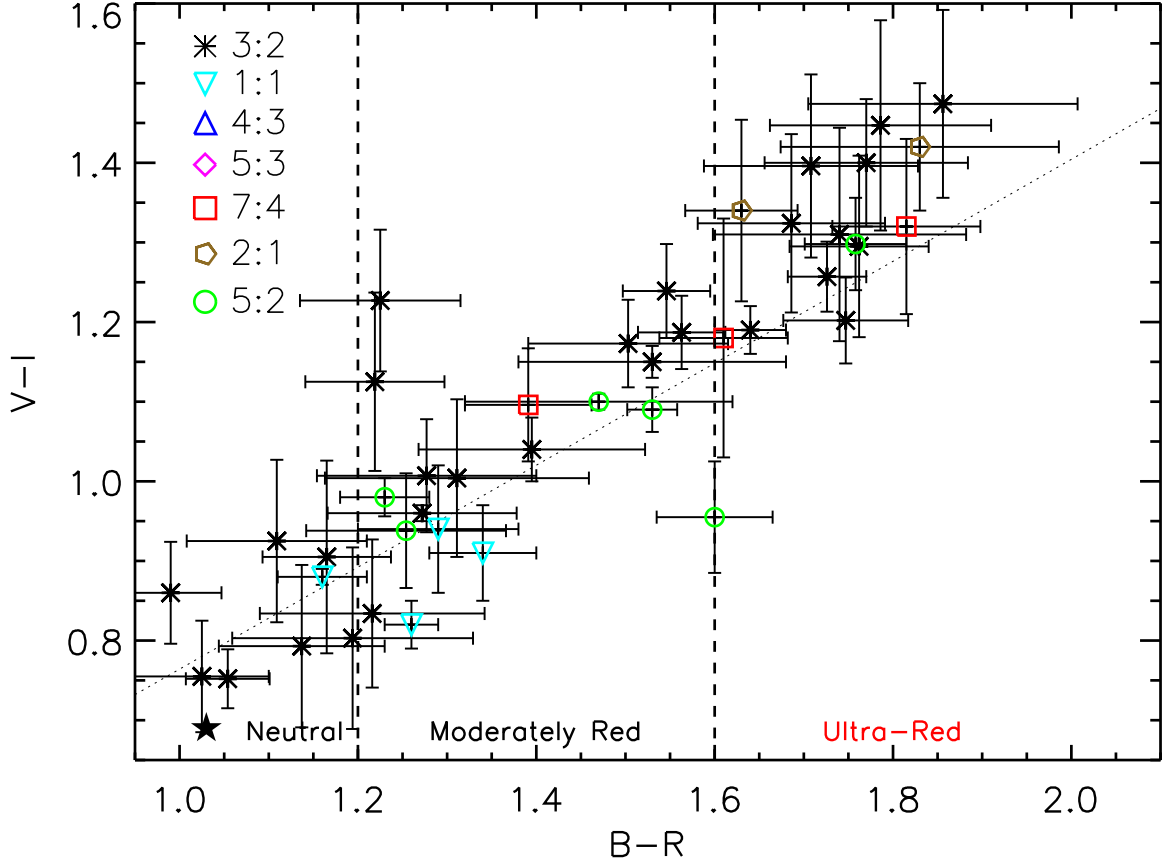}}
\caption{Previously measured BVRI colors of resonant Kuiper Belt
  objects not observed in this work.  Very few resonant objects not in
  the 3:2 population have had previous good color measurements.
  Objects in the literature with error bars significantly larger than
  0.1 magnitudes in color are not used.  Data is from the MBOSS data
  base originally published in Hainaut and Delsanti (2002) as well as
  Luu and Jewitt (1996), Tegler and Romanishin (1998), Tegler and
  Romanishin (2000), Jewitt and Luu (2001), Delsanti et al. (2001),
  Doressoundiram et al. (2002), Tegler et al. (2003), Fornasier et
  al. (2004), Peixinho et al. (2004), Doressoundiram et al. (2005),
  Sheppard and Trujillo (2006), Doressoundiram et al. (2007), DeMeo et
  al. (2009), Santos-Sanz et al. (2009), Sheppard (2010), Snodgrass
  et al. (2010), Romanishin et al. (2010) and Benecchi et al. (2011).}
\label{fig:knownRescolorall}
\end{figure}

The BVRI colors are shown by Neptune resonance membership in Table 3
and Figure~\ref{fig:RescolorallBVRI} for the new observations
presented in this work.  BVRI results for the already well observed
3:2 resonance as well as the few known measurements of other resonance
objects are shown in Figure~\ref{fig:knownRescolorall}.  Ultra-red
color ($S\gtrsim 25$) was coined by Jewitt (2002) and is defined here
as the color 75 percent of all cold classical Kuiper belt objects
have.  Very-red ($S\gtrsim 20$) is defined as the 90 percentile of all
object colors in the cold classical Kuiper belt (Table 5).  The
correlated broadband optical colors of TNOs in
Figure~\ref{fig:RescolorallBVRI} shows they have a nearly linear red
slope in their optical colors.  This near linear optical color slope
has been confirmed through spectroscopy and correlation analysis on
other TNOs (Doressoundiram et al. 2008).  The spectral gradient, S, is
the percent of reddening per 100 nm in wavelength.  Following our
earlier paper Sheppard (2010) we express the spectral gradient as
$S(\lambda_{2} > \lambda_{1}) = (F_{2,V} - F_{1,V}) /(\lambda_{2} -
\lambda_{1})$, where $\lambda_{1}$ and $\lambda_{2}$ are the central
wavelengths of the two filters used for the calculation and $F_{1,V}$
and $F_{2,V}$ are the flux of the object in the two filters normalized
to the V-band filter.  To compute the spectral gradient of the
observed objects in this work with the Sloan filters, we used the g'
and i' measurements.  The g' and i' filters have well-separated
central wavelengths of 481.3 and 773.2 nm, respectively.

It is immediately clear from Table 3 and
Figure~\ref{fig:RescolorallBVRI} that the various Neptune mean motion
resonances have significantly different color distributions.  Just
looking at colors versus resonance occupation shows the Neptune
Trojans are all just slightly red, similar to the Jupiter Trojans
(Karlsson et al. 2009).  The inner 4:3 and outer 5:2 Neptune
resonances are mostly moderately red objects like found in the
scattered disk (Hainaut and Delsanti 2002) and detached disk (Sheppard
2010).  The middle 5:3 and 7:4 Neptune resonances are dominated by
ultra-red objects like found in the low inclination cold classical
belt (Tegler and Romanishin 2000; Trujillo and Brown 2002; Peixinho et
al. 2008).  The 3:2 and 2:1 Neptune resonances have a wide range of
colors from neutral to ultra-red.  Assuming the surface colors of KBOs
are related to their formation location, it is likely the 3:2 and 2:1
Neptune resonances have a greater mix of objects from around the solar
system than the other reservoirs above.  The 5:4 and 12:5 Neptune
resonances have few known objects, but both appear to have significant
numbers of ultra-red objects.  The 3:1 resonance, with only three
measured objects, does not have any known ultra-red members.

\subsection{Size of Resonance Objects}

The two main variables that one may think could significantly bias the
color of objects within a resonance are the object size (Table 3) and
orbital inclination (Table 4).  Except for the 3:2 resonance, almost
all the objects in the other resonances have absolute magnitudes
$m_{R}(1,1,0)\gtrsim 6$ (radii $\lesssim$ 100 km assuming moderate
albedos), and thus are significantly smaller than the dwarf planet
sized objects in the Kuiper Belt (Sheppard et al. 2011).  This means
that even the largest resonance objects are unlikely to have
atmospheres or be strongly differentiated (McKinnon et al. 2008;
Lineweaver and Norman 2010).  This small size is also below the point
where high albedos become prominent as seen on the largest KBOs
(Stansberry et al. 2008; Santos-Sanz et al. 2012).  Size should also
not be a significant factor in biasing the resonant color results as
almost all objects observed in this work have similar absolute
magnitudes ($m_{R}(1,1,0)\sim 7$ mags) independent of which Neptune
resonance they are in (see Table 3).  This is true for all resonances
except maybe the largest objects in the 3:2 resonance.  In the 3:2
resonance, of which Orcus, Pluto and Ixion are dwarf planet sized,
Ixion and Orcus have very different colors.  Orcus is very neutral in
color while Ixion is very red (Doressoundiram et al. 2002; de Bergh et
al. 2005; DeMeo et al. 2009).  Because of the known atmosphere of
Pluto and its ability to significantly alter the surface of an object
(Stern and Trafton 2008), the color of Pluto is not used in this work.

\subsection{Orbital Parameters and Colors of Objects}

\begin{figure}
\epsscale{0.4}
\centerline{\includegraphics[angle=90,totalheight=0.6\textheight]{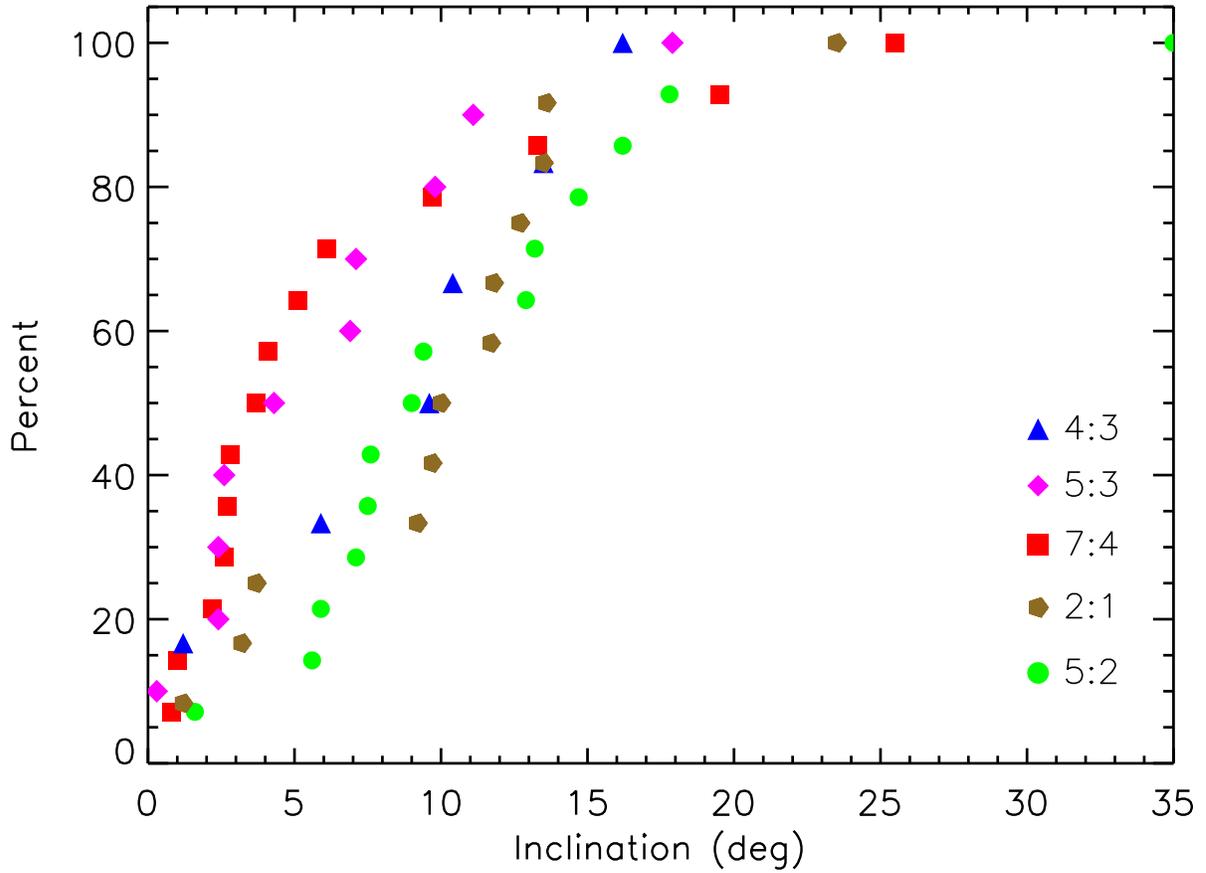}}
\caption{The orbital inclination versus percentage for the main
  Neptune resonances.  The 5:3 and 7:4 resonances have more known low
  inclination objects than the other resonances.  Thus the 5:3 and 7:4
  are not only similar in ultra-red color as the Cold Classical belt
  objects but also have preferentially low inclinations like the Cold
  Classical belt.}
\label{fig:KStestIncl} 
\end{figure}

Inclination has been found to be important for classical KBOs as
objects with inclinations less than about 5 to 10 degrees are
dominated by ultra-red colors (Tegler and Romanishin 2000; Trujillo
and Brown 2002; Stern 2002; Gulbis et al. 2006; Peixinho et al. 2008).
Table 4 shows the orbital elements of the resonant KBOs observed in
this work.  Tables 2 and 3 also show inclinations arranged in
ascending order for each resonance.  Inclination appears to be of some
importance in terms of the number of known objects in each resonance
(Figure~\ref{fig:KStestIncl}).  There are many more known low
inclination objects in the 5:4, 5:3 and 7:4 resonance populations with
few known high inclination objects compared to the other resonances
(see Table 5).  These three resonances also have the highest fraction
of very-red and ultra-red objects (Table 5).

\begin{figure}
\epsscale{0.4}
\centerline{\includegraphics[angle=90,totalheight=0.6\textheight]{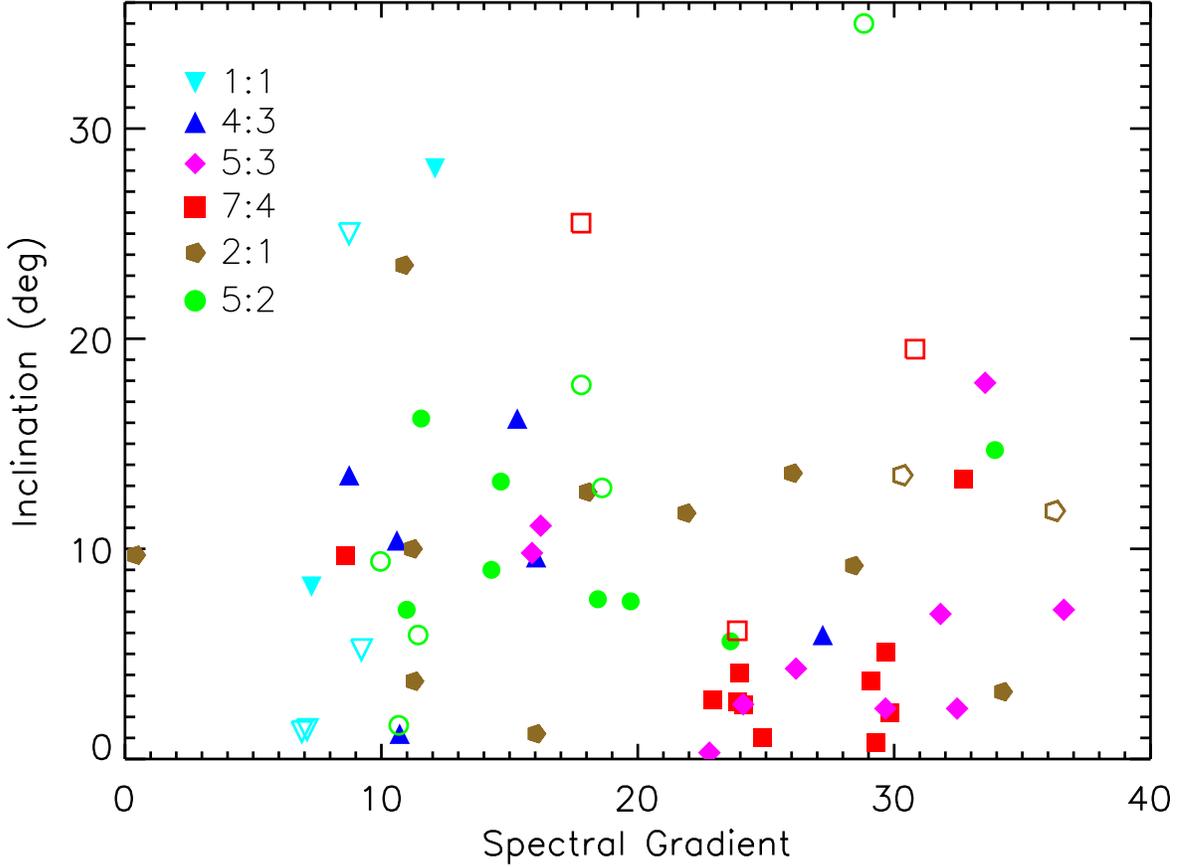}}
\caption{Filled symbols show the spectral gradient versus inclination
  of the main resonances observed in this work.  Open symbols show the
  few measurements found in the literature of resonance objects not
  observed in this work.  There is a lack of low inclination 4:3
  objects and high inclination 5:3 and 7:4 objects, but the few
  observed at these inclinations in these resonances are similar in
  color as the other objects in the particular resonance.  It is
  interesting that the only non ultra-red objects in the 5:3 and 7:4
  resonances have inclinations of about 10 degrees or higher.  It is
  also of note that the only ultra-red object in the 4:3 resonance has
  an inclination less than 10 degrees.  The spectral gradient
  uncertainties have been removed for clarity but can be found in
  Table 2.}
\label{fig:SiRes}
\end{figure}

\begin{figure}
\epsscale{0.4}
\centerline{\includegraphics[angle=90,totalheight=0.6\textheight]{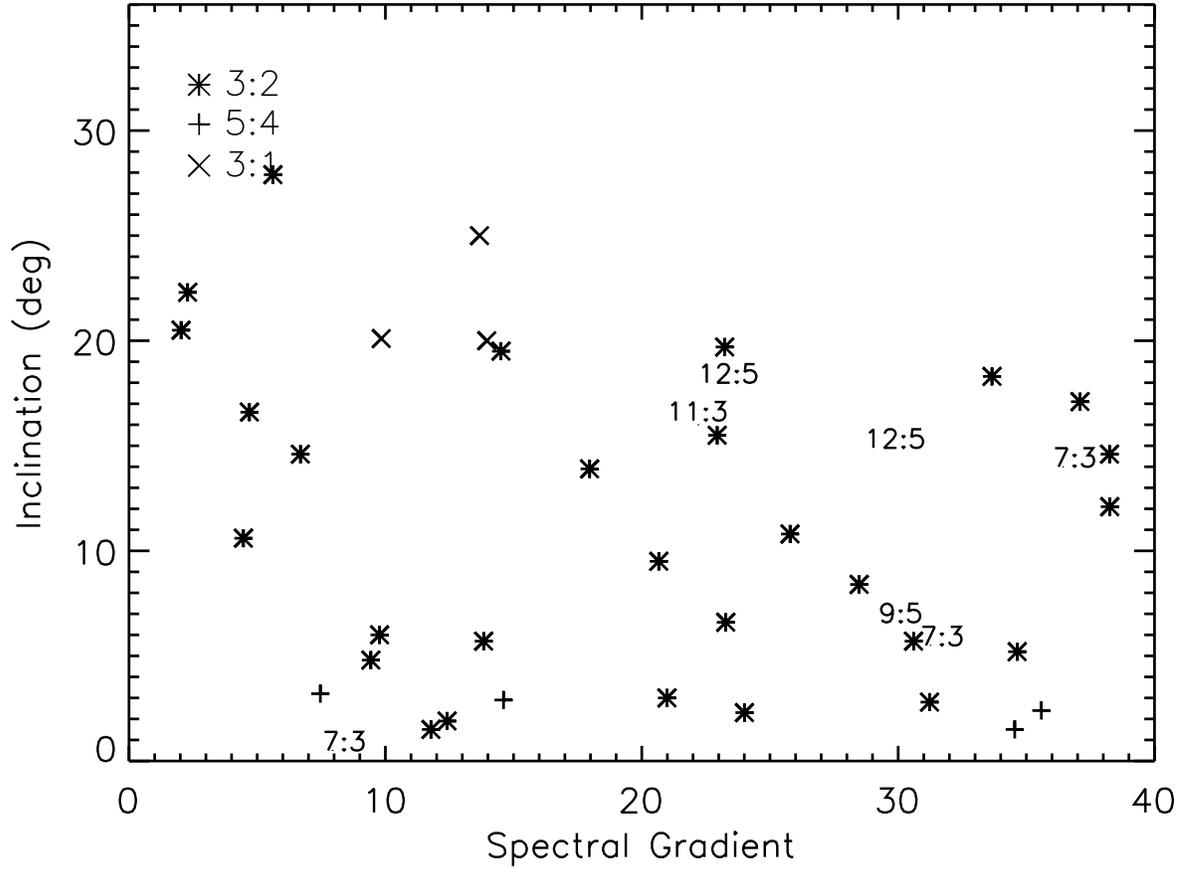}}
\caption{The spectral gradient versus inclination of the 3:2 resonance
  objects and resonances observed with few known members.  The 3:2
  resonance objects (asterisks) appear to have no color inclination
  dependence and have a range of colors from neutral to
  ultra-red.  The 5:4 resonance (pluses) has only low inclination
  members with colors split between moderately red and ultra-red.  The
  3:1 resonance (crosses) shows only high inclination members with
  only moderate redness observed for all three objects.  The 7:3
  resonance has a mix of colors.  Resonances with two or less measured
  members (12:5, 11:3, and 9:5) are shown in the figure by their mean
  motion resonance ratio relative to Neptune.}
\label{fig:knownSiRes}
\end{figure}

The spectral gradients versus inclinations for Neptune resonant
objects are shown in Figures~\ref{fig:SiRes} and~\ref{fig:knownSiRes}.
Even though the 5:3 and 7:4 have mostly low inclination objects, they
also have ultra-red objects at high inclinations.  These ultra-red,
high inclination 5:3 and 7:4 resonant objects could have once been on
low inclination orbits that were excited to higher inclinations
through resonance pumping (Lykawka and Mukai 2005a,2005b; Volk and
Malhotra 2011).  The 5:4 only has very low inclination members with
measured colors (Figure~\ref{fig:knownSiRes}), but even with so few of
them known, it appears to have a mix of colors and not be as dominated
by ultra-red material as the 5:3 and 7:4 resonances (Table 5).  In
contrast, the 12:5 only has known high inclination members, of which
both have ultra-red colors (Figure~\ref{fig:knownSiRes}).  The 3:1 has
no ultra-red objects but all measured 3:1 objects are of high
inclination (Figure~\ref{fig:knownSiRes}).

The only non ultra-red objects in the 5:3 and 7:4 resonances have
inclinations greater than about 10 degrees, suggesting 10 degrees is
the place to distinguish between the low inclination cold classical
and high inclination classical belt populations.  All objects in the
5:3 and 7:4 resonances below about 10 degrees are ultra-red.  More low
inclination 4:3 and high inclination 5:3 and 7:4 resonant objects need
to be found and measured for colors to determine how statistically
significant inclination is.  As of the time of this writing, the known
numbers of objects at some inclinations are very low in these
resonances.  It seems that the 5:4 and 3:1 resonant objects may only
occupy low and high inclination orbits, respectively.

\begin{figure}
\epsscale{0.4}
\centerline{\includegraphics[angle=90,totalheight=0.6\textheight]{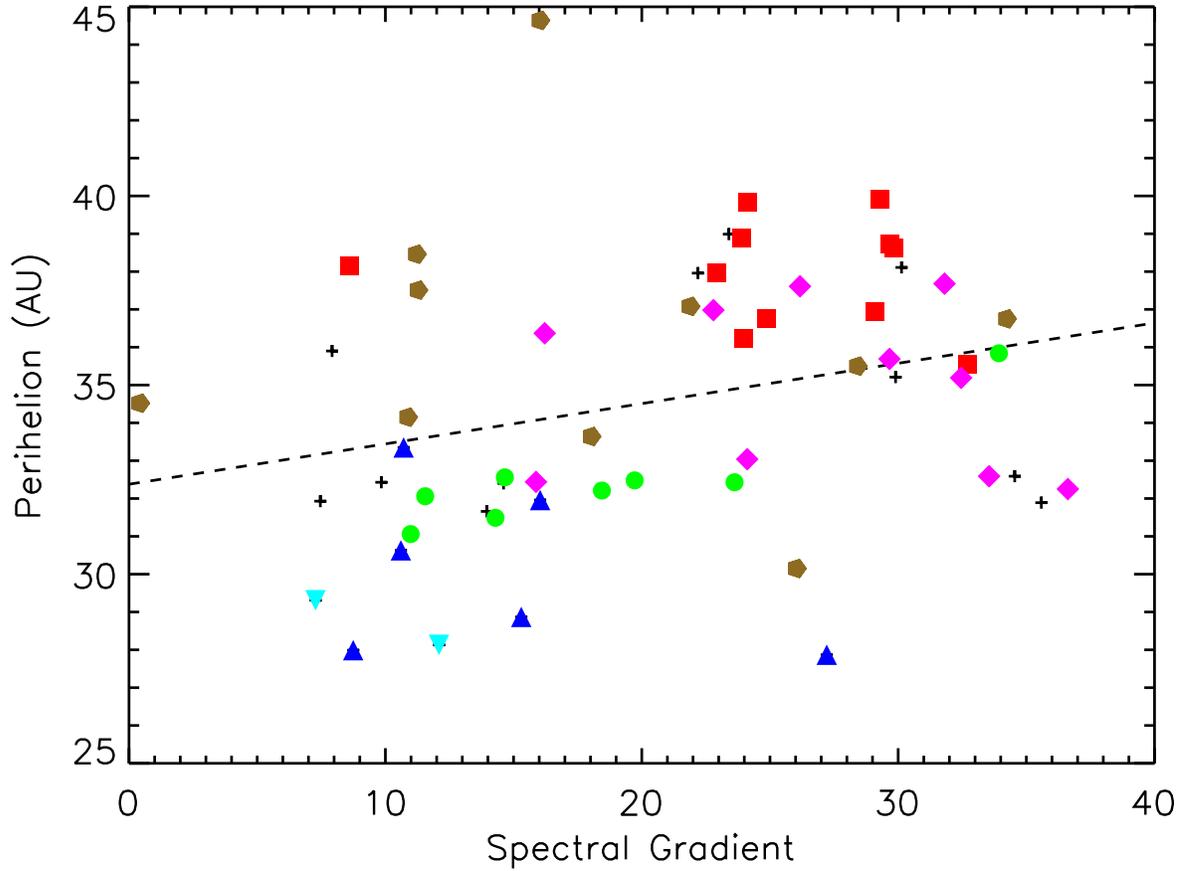}}
\caption{Same as Figure~\ref{fig:SiRes} except comparing the spectral
  gradient of the various resonance objects to their perihelion
  distances.  The dashed line is the fit to all the data points.
  There is a moderate correlation that the higher perihelion objects
  have redder surfaces, but it is only significant at about the $97\%$
  confidence level.  The non major less populated resonant objects are
  shown as plus signs.}
\label{fig:SqRes}
\end{figure}

\begin{figure}
\epsscale{0.4}
\centerline{\includegraphics[angle=90,totalheight=0.6\textheight]{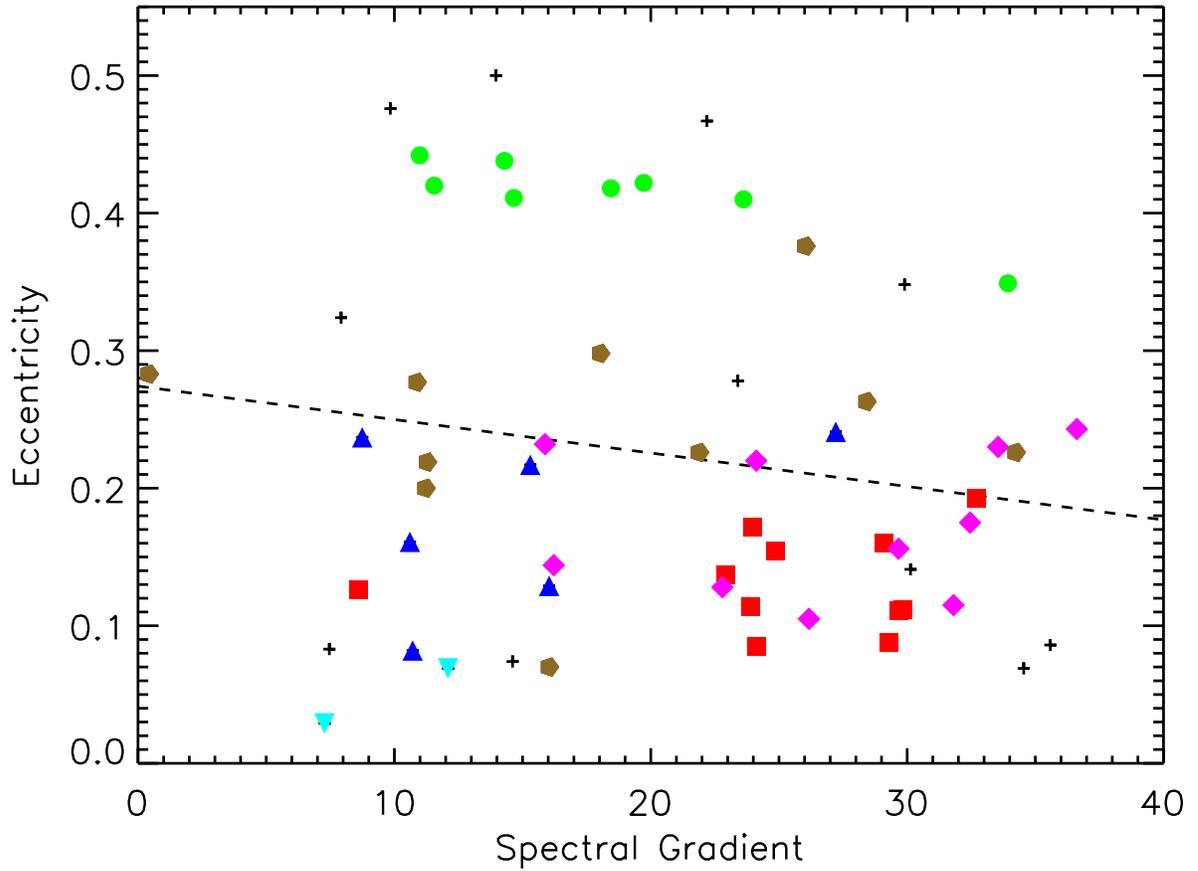}}
\caption{Same as Figure~\ref{fig:SiRes} except comparing the spectral
  gradient of the various resonance objects to their eccentricities.
  There is a slight trend that lower eccentricity objects tend to be
  redder in color.  The non major less populated resonant objects are
  shown as plus signs.}
\label{fig:SeRes}
\end{figure}

\begin{figure}
\epsscale{0.4}
\centerline{\includegraphics[angle=90,totalheight=0.6\textheight]{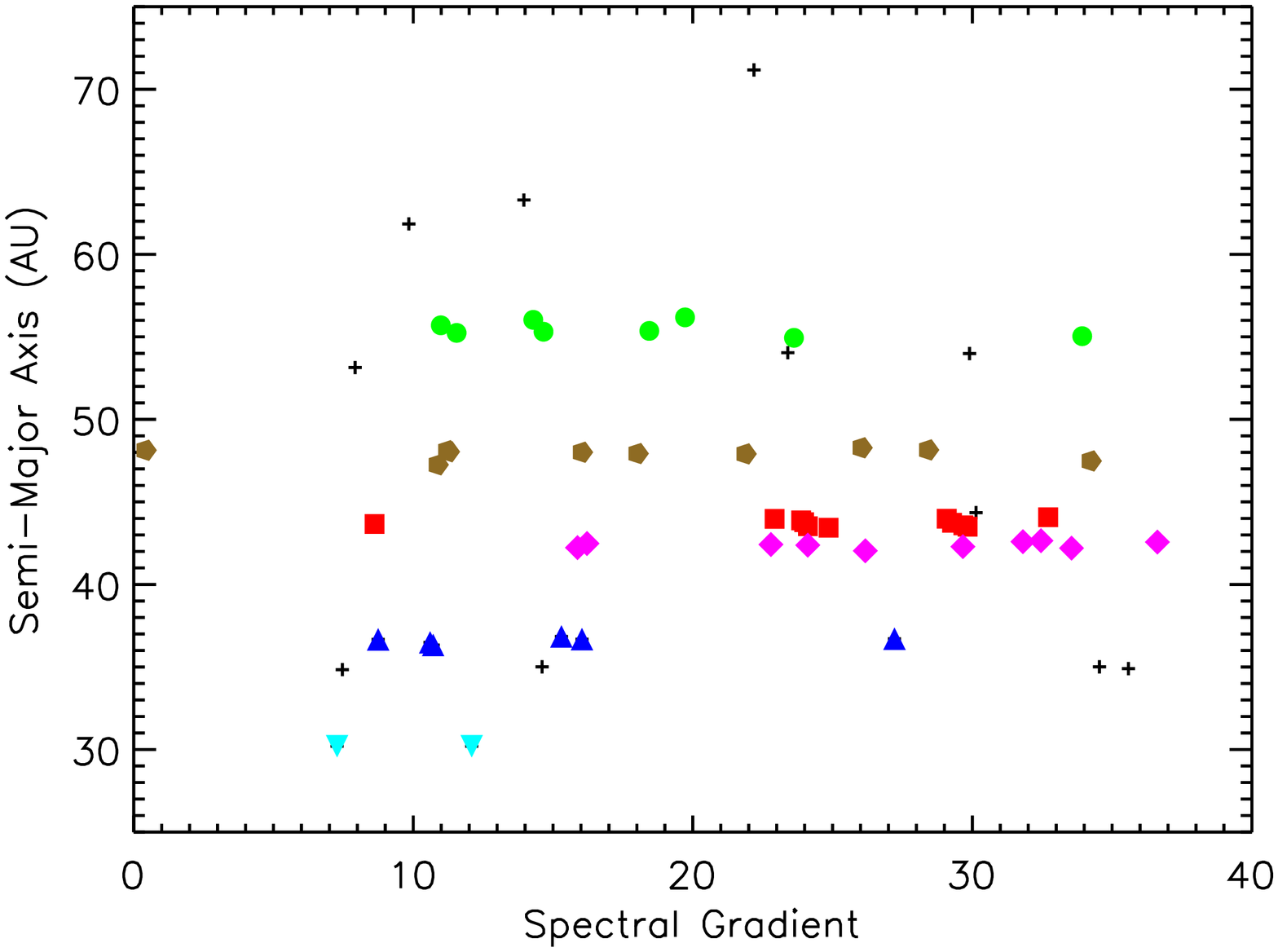}}
\caption{Same as Figure~\ref{fig:SiRes} except comparing the spectral
  gradient of the various resonance objects to their semi-major axes.
  The non major less populated resonant objects are shown as plus
  signs.}
  \label{fig:SaRes}
\end{figure}

Tegler and Romanishin (2000) suggest that high perihelion distance and
low eccentricity are also important quantities for the ultra-red
material in the classical Kuiper Belt.  Figures~\ref{fig:SqRes}
to~\ref{fig:SaRes} compare the spectral gradient of the resonance
objects to their perihelion distances, eccentricities and semi-major
axes.  There are no obvious strong correlations, but there is a
moderate correlation at about the $97\%$ confidence level (Pearson
coefficient of 0.28 with a sample of 58 objects) that resonant objects
with high perihelion distances are redder (Figure~\ref{fig:SqRes}).
These object color results versus the dynamics of the objects are
similar to the cold classical object results for the location of
ultra-red material.  There is still a strong debate as to why objects
with higher perihelia would have redder colors and thus likely
different surface compositions than objects with lower perihelia.
This color difference could be related to the retention and/or
irradiation of very volatile ices such as Ammonia, Methane or Methanol
at large heliocentric distances (Schaller and Brown 2007; Grundy 2009;
Brown et al. 2011; Merlin et al. 2012).  For example, Methane's
condensation temperature is around 40K, which in our Solar System
happens around 48 AU (Youdin and Kenyon 2012).  This is a similar
distance as the cold classical Kuiper belt and 5:3 and 7:4 resonances.

\section{Discussion}

\subsection{Resonant Colors Compared}

\begin{figure}
\epsscale{0.4}
\centerline{\includegraphics[angle=90,totalheight=0.6\textheight]{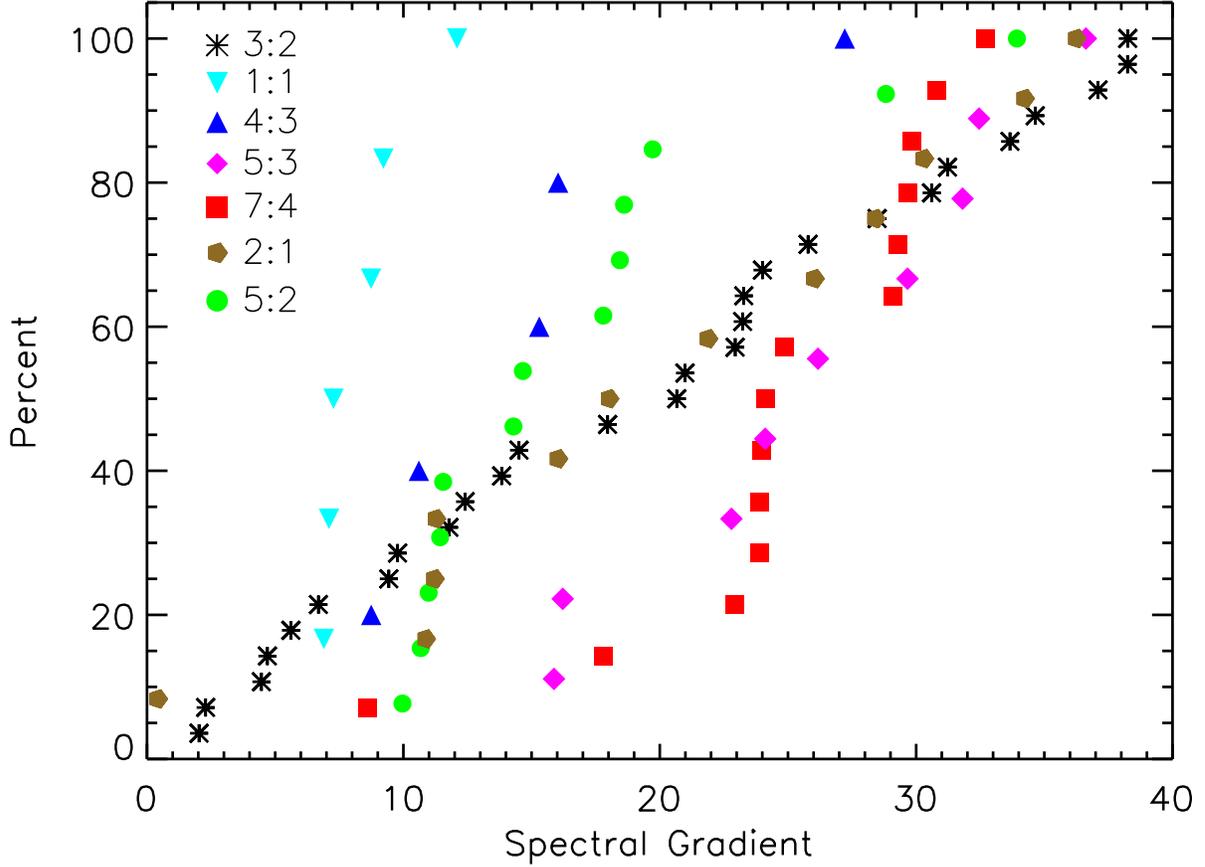}}
\caption{The Kolmogorov-Smirnov test (K-S test) plotted for the
  Neptune Trojans 1:1 (cyan upside down triangles), 4:3 (blue
  triangles), 5:3 (purple diamonds), 7:4 (red squares), 2:1 (brown
  pentagons), 5:2 (green circles) and 3:2 (black asterisks).  The
  vertical axis shows the cumulative spectral gradient, S, for the
  objects, where ultra-red is $S\gtrsim 25$.  The Neptune Trojans
  (1:1) are the most neutral in color and uniform of all the
  resonances.  The 4:3 and 5:2 are mostly only moderately red in
  color, though the 4:3 resonance has very few low inclination ($i<10$
  deg) objects with measured colors.  The 5:3 and 7:4 resonance
  objects are mostly ultra-red though neither resonance has many high
  inclination objects with measured colors.  The 2:1 and 3:2
  resonances seem to have a mix of all object colors and are not
  dependent on inclination.}
\label{fig:KStestRes} 
\end{figure}

To compare the various resonant color populations directly, the
Kolmogorov-Smirnov test and Student's t-test were calculated using the
known color data (Student 1908; Kolmogorov 1933; Smirnov 1948).
Figure~\ref{fig:KStestRes} displays the data graphically while the
results are shown in Table 6.  The 5:3 and 7:4 resonances both have a
very similar ultra-red color distribution and thus could be drawn from
the same parent population.  The 4:3 and 5:2 Neptune resonances also
have similar color distributions, being mostly moderately red objects.
It is possible that the 4:3 and 5:2 populations are from the same
parent population.  As shown in Table 6, the 5:3/7:4 resonant objects
can be rejected as having the same common parent population as the
4:3/5:2 resonant objects at about the $99\%$ confidence level,
assuming the currently observed KBO colors are from the original
formation location within the solar system.  The 2:1 and 3:2 resonant
colors both cover a much wider color distribution than the other
resonances and could be drawn from the same parent population.  The
2:1 and 3:2 seem to have a mix of object colors and may represent many
different originally separate parent populations.  The 3:2 also
appears to have the only significant population of neutral colored
objects within the resonant populations.  The Neptune Trojans seem to
be unique and do not appear to be drawn from the same parent
population as any of the other Neptune resonances.

\subsection{Resonant Colors Compared to Other KBO Classes}

\begin{figure}
\epsscale{0.4}
\centerline{\includegraphics[angle=90,totalheight=0.6\textheight]{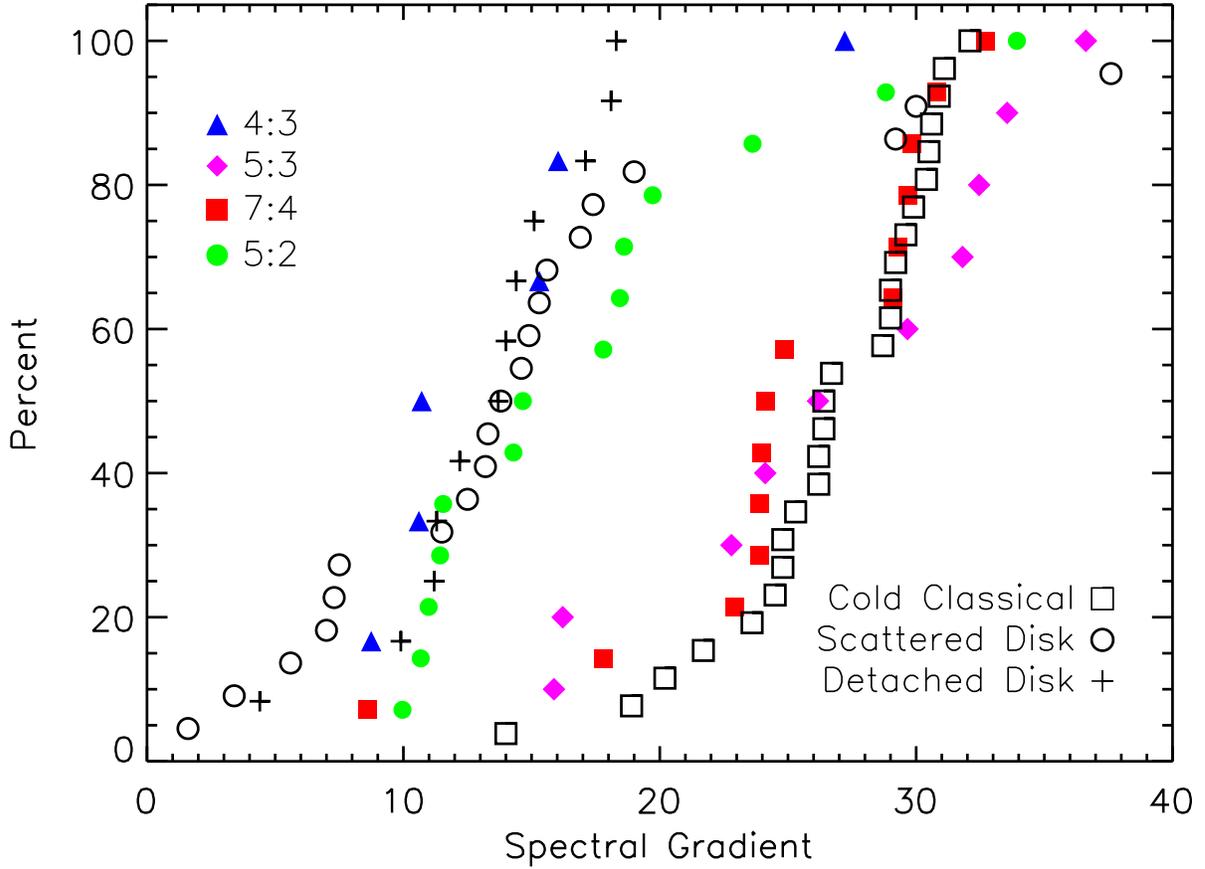}}
\caption{Same as Figure~\ref{fig:KStestRes} except the Scattered Disk
  (open circles), Detached Disk (pluses), and Cold Classical belt
  objects (open squares) are shown to compare to the 4:3, 5:3, 7:4 and
  5:2 resonances.  The 1:1, 3:2 and 2:1 resonances have been removed
  for clarity.  The 4:3 and 5:2 resonances appear to be very similar
  in color distribution as the scattered disk and detached disk
  objects while the ultra-red dominated 5:3 and 7:4 resonant
  populations are very similar in color distribution to the Cold
  Classical Kuiper belt.}
\label{fig:KStestKnown} 
\end{figure}

Figure~\ref{fig:KStestKnown} further compares the various resonant
color populations to the main Kuiper Belt reservoirs defined in the
introduction such as the cold classical Kuiper belt, scattered disk
and detached disk.  As shown in Table 6, the ultra-red dominated 5:3
and 7:4 Neptune resonances appear to be drawn from the same parent
population as the cold classical belt objects.  This suggests that the
5:3 and 7:4 resonant objects are just a continuation of the cold
classical Kuiper Belt.  If true, high inclination objects may not be
expected to be a significant fraction of the 5:3 and 7:4 populations,
though some orbit inclination modification is expected for objects in
and around the 5:3 and 7:4 resonances (Lykawka and Mukai 2005a,2005b;
Volk and Malhotra 2011).

In contrast, the 4:3 and 5:2 resonances may have similar origins as
the scattered disk, detached disk and/or high inclination classical
belt objects based on colors.  Though low number statistics, the 3:1
resonance is probably similar to the 4:3 and 5:2 resonances.  A
similar origin for the 4:3, 5:2 and 3:1 resonances would not only
explain their similar colors but also these populations lack of low
inclination members.  The low number of ultra-red type objects in the
4:3, 5:2 and 3:1 suggest these objects might have once been much
closer to the Sun, where as discussed in section 3.2, highly volatile
ice rich ultra-red material on their surfaces could be destroyed.
These objects could have been later scattered and captured into their
current distant, mostly high inclination, resonant orbits.

\subsection{Neptune's Migration History}

Lykawka and Mukai (2005b) determined that the 2:1 resonance should be
more heavily populated than the 5:3 or 7:4 based on numerical
simulations of objects in the classical belt and Neptune's orbital
history.  Murray-Clay and Schlichting (2011) suggest the resonant
populations should have a low inclination ``cold'' component to their
populations if Neptune had a slow and smooth migration.  One might
also expect these resonant ``cold'' components to be dominated by
ultra-red material like found in the cold classical Kuiper belt (24 of
26 objects or $92\%$ are very red or redder in color, see Table 5).
As discussed above and shown in Table 5, very red colors dominate the
5:3 and 7:4 resonances and this might also be true for the sparsely
known inner 5:4 Neptune resonance.  The inner 3:2 resonance may also
show a cold component as this resonance has 8 of 13 ($67\%$) known
objects with low inclination as being very red or redder.  Though this
is statistically the same amount of very red material at higher
inclinations in the 3:2 resonance, putting into question if the 3:2
has a true cold component or if the low inclination objects are just
simply a continuation of the 3:2 resonance as a whole.  High fractions
of very red objects at low inclinations do not appear to be true in
any of the outer resonances observed to date.  The 2:1 and 5:2 Neptune
resonances both have a few known objects below 10 degrees inclination
and thus could have a cold component, but these low inclination
objects are not preferentially ultra-red.  On the contrary, only 3 of
the 11 ($27\%$) objects known with inclinations less than 10 degrees
in the 2:1 and 5:2 resonances have very-red or ultra-red colors.  This
small very red or redder fraction is also true for objects in the 2:1
and 5:2 resonances using the statistically smaller samples of objects
with inclinations less than 8 and 5 degrees, 2 of 7 ($29\%$) and 1 of
3 ($33\%$), respectively.  Because the cold classical objects are so
dominated by very red material ($92\%$, Table 5), simple probability
statistics (for example, the probability of 2 of the 3 known outer
resonant objects with inclination less than 5 degrees observed for
colors having only moderately red color if drawn from the cold
classical objects color distribution would be $2/26 \times 1/25$) give
about a 3 sigma result that the low inclination objects observed in
the 2:1 and 5:2 resonances have a less red color distribution than the
cold classical Kuiper belt objects.  This suggests if there is a cold
component to the outer resonances, it is composed of different objects
or that the objects had significantly different environmental
histories than the cold components of the inner and middle resonances.

In the slow smooth migration model, the outer Neptune resonances would
have swept gently through the cold classical Kuiper belt region.
Assuming the cold classical objects were fully formed, the slow smooth
migration model would predict that many of the ultra-red cold
classicals would have been captured into these outer resonances during
this time.  Assuming the colors of objects captured into the outer
resonances are the same today, the significant differences in the
various resonant population colors and the small number of ultra-red
objects in the low inclination populations of the outer resonances
suggests Neptune did not experience a significant slow smooth
migration phase.  Neptune likely had a much more chaotic migration
history with a large eccentricity just after formation and scattering
allowing many various small bodies to be captured in the outer Neptune
resonances (Levison et al 2008).  The absence of any obvious cold
component in the outer resonances also agrees with the Hahn and
Malhotra (2005) result that Neptune migration likely occurred after
the Kuiper Belt was already dynamically stirred-up.

If the 5:3 and 7:4 resonant populations share a common origin with
objects in the cold classical belt, it is expected that the 5:3 and
7:4 should have a high number of equal-sized, ultra-red binaries.
This is because the cold classical belt has been found to have a
binary fraction of about $\sim 30\%$ while the other types of Kuiper
Belt objects have only about a $5\%$ binary fraction (Noll et
al. 2008).  To date, the only known ultra-red, equal-sized binary
found outside of the cold classical belt is 2007 TY430 (Sheppard et
al. 2012).  2007 TY430 likely became ``stuck'' in the 3:2 resonance
population after escaping from the cold classical region (Lykawka and
Mukai 2006).


\section{Summary}

Fifty-eight Neptune mean motion resonance objects were observed for
their optical surface colors.  The various Neptune mean motion
resonances were found to have significantly different object surface
color distributions.  This indicates vastly different origins and
evolutions for the objects in resonance with Neptune.  Ultra-red color
($S\gtrsim 25$, g'-i'$\gtrsim 1.2$ mags, B-I$\gtrsim 2.2$ mags) is
defined here as the color 75 percent of all cold classical Kuiper belt
objects have.  Very-red color ($S\gtrsim 20$) is defined as the 90
percentile of all object colors in the cold classical Kuiper belt.

1. The 5:3 and 7:4 Neptune resonances are composed of mostly ultra-red
objects.  The few known high inclination objects in these resonances
are also ultra-red in color.  The colors of the 5:3 and 7:4 resonant
objects are statistically identical to the low inclination cold
classical belt colors.  Thus the 5:3 and 7:4 resonant objects are
likely just an extension of the cold classical belt.  If true, the 5:3
and 7:4 objects should have a significant number of ultra-red, equal
sized binaries as have been found in the cold classical belt.  The
only non very-red colored objects in the 5:3 and 7:4 resonances all
have inclinations near 10 degrees, suggesting this is the inclination
region that separates the low inclination cold classical and high
inclination classical belt populations.

2. The inner 4:3 and outer 5:2 Neptune resonances have few ultra-red
objects and are composed of mostly moderately red objects.  These two
resonances, along with the sparsely sampled 3:1 Neptune resonance,
appear to have a similar color distribution as the scattered disk
(including the high inclination, moderate eccentricity classical
objects) and detached disk.  If these populations all have a similar
origin, it would also explain the abundance of high inclination
objects and lack of low inclination objects found in these three
Neptune resonances.  

3. In contrast to the narrow color distributions found for the above
resonances, the 2:1 and 3:2 resonances have a very wide color
distribution.  This indicates these two resonances likely captured
objects that formed in many different places within the solar system,
assuming the colors are uniquely associated with origin radius.

4. The Neptune Trojans appear to be unique in surface color for outer
solar system objects as these objects are only slightly red.  They are
very similar to the Jupiter Trojans in color.

5.  There is a moderate correlation that the higher perihelion
resonant objects have redder surfaces, but it is only significant at
about the $97\%$ confidence level.

6. If there are low inclination ``cold'' components of the inner 3:2
and 5:4 resonances, they could be composed of a large fraction of
ultra-red objects, like found for the middle 5:3 and 7:4 resonances.
This is because there are many objects with low inclinations in the
3:2 and 5:4 with very red colors.  With the limited color data to
date, the outer 2:1 and 5:2 resonances do not show a high fraction of
ultra-red objects at low inclinations.  This suggests if there is a
cold component in the outer resonances, the surfaces of the objects
are different than the cold components of the middle Neptune
resonances as well as the cold classicals.  If true, this makes it
unlikely Neptune had a significant slow smooth migration phase in its
past since these outer resonances would be expected to have a
significant cold component similar to the ultra-red objects in the
cold classical Kuiper belt that the resonances would have swept
through.

\section*{Acknowledgments}
The author thanks C. Trujillo and S. Benecchi for helpful comments and
suggestions while writing this manuscript.  This paper includes data
gathered with the 6.5 meter Magellan Telescopes located at Las
Campanas Observatory, Chile.  S. S. was partially supported by the
National Aeronautics and Space Administration through the NASA
Astrobiology Institute (NAI) under Cooperative Agreement
No. NNA04CC09A issued to the Carnegie Institution of Washington.

\newpage

%
%
%
%



\begin{center}

\end{center}


\newpage

\newpage

\newpage

\newpage

\newpage

\newpage

\newpage

\newpage

\newpage

\newpage

\newpage

\newpage

\newpage

\end{document}